\documentclass[twocolumn,showpacs,amsmath,amssymb,pra,nofootinbib]{revtex4}

\usepackage{graphicx}
\usepackage{dcolumn}
\usepackage[dvips]{color}
\usepackage{times,mathptm,amssymb}
\usepackage[ps2pdf,colorlinks]{hyperref}

\newcommand{\cblu}[1]{{\color{blue}{#1}}}

\newcommand{\cred}[1]{{\color{myred}{#1}}}

\definecolor{gold}{rgb}{0.85,.66,0}
\definecolor{plum}{rgb}{0.45,0,.66}
\definecolor{bgn}{rgb}{0,.85,.46}  
\definecolor{myred}{rgb}{1,.1,.16}

\newcommand{\bmx}{\left[ \begin{array}}
\newcommand{\emx}{\end{array} \right]}
\newcommand{\be}{\begin{equation}}
\newcommand{\ee}{\end{equation}}
\newcommand{\bea}{\begin{eqnarray}}
\newcommand{\eea}{\end{eqnarray}}

\newcommand{\ket}[1]{|#1\rangle}
\newcommand{\bra}[1]{\langle#1|}

\newcommand{\bl}{\begin{itemize}}
\newcommand{\el}{\end{itemize}}
\newcommand{\ii}{\item}

    {
    \smallskip
    \refstepcounter{theorem}
    \noindent
    {\bf Example \Alph{section}.\arabic{theorem}} \ \ }
    {\hspace*{\fill}{$\Diamond$}
    \smallskip}

    {
    \smallskip
    \refstepcounter{theorem}
    \noindent
    {\bf Remark \Alph{section}.\arabic{theorem}} \ \ }
    {\hspace*{\fill}{$\Diamond$}
    \smallskip}

    {
    \smallskip
    \refstepcounter{theorem}
    \noindent
    {\bf Definition \Alph{section}.\arabic{theorem}} \ \ }
    {\hspace*{\fill}{\ }
    \smallskip}

    {
    \smallskip
    \refstepcounter{theorem}
    \noindent
    {\bf Example \arabic{section}.\arabic{theorem}} \ \ }
    {\hspace*{\fill}{$\Diamond$}
    \smallskip}

    {
    \smallskip
    \refstepcounter{theorem}
    \noindent
    {\bf Remark \arabic{section}.\arabic{theorem}} \ \ }
    {\hspace*{\fill}{$\Diamond$}
    \smallskip}

    {
    \smallskip
    \refstepcounter{theorem}
    \noindent
    {\bf Definition \arabic{section}.\arabic{theorem}} \ \ }
    {\hspace*{\fill}{\ }
    \smallskip}

    {
    \smallskip
    \refstepcounter{theorem}
    \noindent
    {\bf Definition \Alph{section}.\arabic{theorem}} \ \ }
    {\hspace*{\fill}{\ }
    \smallskip}

    {
    \smallskip
    \refstepcounter{theorem}
    \noindent
    {\bf Scholium \arabic{section}.\arabic{theorem}} \it \ \ }
    {\hspace*{\fill}{\ }
    \smallskip}

\newenvironment{proof}[1][]
    {
    \noindent
    {\bf Proof{#1}:  }
    }
    {\hspace*{\fill}{$\Box$}\smallskip}

    {
    \noindent
    {\bf Sketch{#1}:  }
    }
    {\hspace*{\fill}{$\Box$}\smallskip}

    {
    \noindent
    {\bf Reference:  }
    }
    {\hspace*{\fill}{$\odot$}\smallskip}

\newtheorem{theorem}{Theorem}[section]
\newtheorem{proposition}[theorem]{Proposition}

\begin{document}
\preprint{APS}
\bibliographystyle{apsrev}

\title{Parallelism for Quantum Computation with Qudits}

\author{Dianne P. O'Leary$^{1,3}$}\email{oleary@cs.umd.edu}
\author{Gavin K. Brennen$^2$}\email{gavin.brennen@uibk.ac.at}
\author{Stephen S. Bullock$^3$} \email{ssbullo@super.org}

\affiliation{
$^1$ University of Maryland, Department of Computer Science
and Institute for Advanced Computer Studies, College Park,
Maryland 20742 \\
and National Institute of Standards and Technology, Mathematical and
Computational Sciences Division, Gaithersburg, MD  20899 USA\\
$^2$ Institute for Quantum Optics and Quantum Information of the 
Austrian Academy of Sciences, A-6020, Innsbruck, Austria \\
$^3$ IDA Center for Computing Sciences,
17100 Science Drive, Bowie, MD 20715-4300 USA \\
}
\setlength{\unitlength}{0.008in} 
\date{\today}
\begin{abstract} 
Robust quantum computation with $d$-level quantum systems 
(qudits) poses two requirements:
fast, parallel quantum gates and high fidelity two-qudit gates. 
We first describe how to implement parallel single qudit operations.
  It is by now well known that
any single-qudit unitary can be decomposed into a sequence of 
Givens rotations on two-dimensional subspaces of the qudit state space.  
Using a  coupling graph to represent physically allowed couplings 
between pairs of 
qudit states, we then show that the logical depth of the parallel gate 
sequence is equal to the height of an associated tree.  
The implementation of a given unitary can then optimize 
the tradeoff between gate time and resources used.
These ideas are illustrated for qudits encoded in the ground hyperfine states
of the atomic alkalies $^{87}$Rb
and $^{133}$Cs.  Second, we provide a protocol for 
implementing parallelized non-local 
two-qudit gates
using the assistance of entangled qubit pairs.  Because the 
entangled qubits can be prepared non-deterministically, this 
offers the possibility of high fidelity two-qudit gates. 
\end{abstract}

\pacs{03.67.Lx}

\keywords{qudit, universal quantum computation}

\maketitle

\section{Introduction}
 
Quantum computation requires the ability to process 
quantum data on a time scale that is small compared to the error rate 
induced by environmental interactions (decoherence).  
Robust computation results when the rate of error in the 
control operations and the rate of decoherence is below some threshold  
independent of the size of the computational register.  
The threshold theorem implies such rates exist,
but it assumes arbitrary connectivity 
between subsystems as well as the ability to implement the 
control operations with a high degree of parallelism \cite{Preskill}.
Quantum computer architectures, therefore, should be designed to 
support parallel gate operations and measurements.   
At the software level some work has been done regarding parallel computation 
with qubits.  For example, certain quantum algorithms such as the 
quantum Fourier transform can be parallelized \cite{Moore:98}, 
and there are techniques to 
compress the logical depth of a quantum circuit on qubits using the 
commutativity of gates in the Clifford group \cite{Raussendorf:02}.  
Further, by using distributed entanglement resources, some frequently 
used control operations can be parallelized \cite{Anocha:04}.

This work concerns parallel unitary operations on qudits, i.e. $d$ level
systems where typically $d>2$.  There are several reasons for 
considering such systems.  Many physical candidates for 
quantum computation with qubits work by encoding 
in a subspace of a system with many more accessible levels.  Control
over all the levels is important for state preparation, simulating quantum 
processes, and measurement.  In particular, encoding in decoherence-free
subspaces usually involves control over multiple distinguishable
states.  Additionally, for small quantum computations, a fixed unitary $U\in U(d)$
 for $d$ small but larger than $2$,
 can often be implemented 
with higher fidelity in a single qudit rather than by simulation with two-qubit gates.
Further, at the level of tensor structures, some quantum processing 
may be more efficient with qudits, e.g. the Fourier transform over
an abelian group whose order is not divisible by two \cite{Hoyer}.  
It is straightforward to show that na{\"\i}ve
qubit emulation of qudits is inefficient \cite{BOBII}.

Fast single-qudit gate times are important in order 
to implement quantum error correction before
errors accumulate \cite{Gottesman99}.  In Section \ref{singlepll}
we derive \emph{parallel} 
implementations of general 
one-qudit unitary gates, where the quantum one-qudit gate
library is restricted to a small set of couplings 
between two-dimensional subspaces (Givens rotations).  The choice of
this Givens library of one-qudit gates reflects standard coupling diagrams,
i.e. the particular rotations obey selection rules in the physical
system that encodes the qudit.   Prior work considered minimum-gate 
circuits for such
generalized coupling diagrams but did not further optimize these
circuits in terms of depth \cite{selectionQR}.  Parallelism 
is possible because
quantum gates on disjoint subspaces can be applied simultaneously,
at the expense of additional control resources.
Our method is particularly helpful for experimental 
implementations because 
it can be applied to a large class of systems with 
different allowed physical
couplings.  We provide examples for qudit control 
with ground electronic
hyperfine levels of $^{87}${\tt Rb}
and $^{133}${\tt Cs} and show that it is possible to achieve
impressive speed-up with these systems using 
three pairs of control fields.
 
Further, in Section \ref{remote}
we obtain depth-optimized
(parallel) implementations of non-local two-qudit gates.   
Specifically, we describe how these operations, which 
generically require $O(d^3)$ elementary two-qudit gates, can 
be parallelized to 
depth $O(d^2)$ using $O(d^3)$ maximally entangled {\it qubit} 
pairs (e-bits).  While the protocol is not optimized 
in terms of e-bits consumed, it is a step forward to the goal 
of high fidelity two-qudit gates.  The qubit resources can
be chosen to be ancillary degrees of freedom of the particle encoding the 
qudit. Thus they can be prepared in entangled pairs 
non-deterministically and purified before the non-local gate is implemented.  
 
A third aspect of parallelism \cite{Moore:98} 
involves reducing the logical depth of a circuit by judicious 
grouping of single- and two-particle gates that can be performed at 
the same time step, assuming connectivity of the particles.   
This is roughly analogous to classic circuit layouts and will
not be considered here.

\section{Parallelism in state synthesis and unitary transformation
for a single qudit}
\label{singlepll}

In typical physical systems encoding a single qudit, arbitrary 
couplings are not allowed. 
Whereas we can represent any unitary $U\in U(d)$ as an 
operator generated from 
an appropriate set of Hamiltonians, viz.
$U=e^{-i \sum_{j=0}^{d^2} t_j h_j}$ where $t_j\in \mathbb{R}$ and
$\sqrt{-1} h_j \in\mathfrak{u}(d)$
with $h_j=h_j^{\dagger}$, it is generally not 
possible to turn on all the 
couplings $h_j$ at the same time.  It is a problem of quantum 
control to determine how 
to simulate a single-qudit unitary using a sequence of available couplings. 

Because quantum computations need only be simulated up to a global phase, 
we restrict ourselves to implementations of a generic unitary $U\in SU(d)$.  
One way to a implement $U$ is by a covering with 
gates generated by the $\mathfrak{su}(2)$
subalgebras $\mathfrak{g}_{j,k}$ acting on the subspaces spanned by
the state pairs $(\ket{k},\ket{j})$: 
\begin{equation}
\begin{array}{lcll}
\mathfrak{g}_{j,k} &  = & \big\{ \; i \lambda_{j,k}^{x,y,z} \; ; & 
\lambda^x_{j,k}=\ket{j}\bra{k}+\ket{k}\bra{j},  \\
& & & \lambda^y_{j,k}=-i(\ket{j}\bra{k}-\ket{k}\bra{j}), \\ 
& & & \lambda^z_{j,k}=\ket{j}\bra{j}-\ket{k}\bra{k} \quad \quad \big\} \\
\end{array}
\end{equation}
This is realized by a $QR$ decomposition of the {\it inverse} 
unitary into a product
of unitary (Givens) rotation matrices that reduce it to diagonal
form $D^{\dagger}$:
\begin{equation}
\label{eq:QR}
D^{\dagger}=\Big[\prod_{\ell=1}^{d(d-1)/2}G_{j_{\ell}k_{\ell}}\Big]U^{\dagger}.
\end{equation}
Here, each Givens rotation can be chosen to 
be a function of two real parameters only:
\begin{equation}
\label{eq:givens}
G_{jk}(\gamma,\phi)=
e^{-i\gamma(\cos\phi \lambda^x_{j,k} - \sin\phi \lambda^y_{j,k})}.
\end{equation}
Typically, parameters are chosen so that 
consecutive Givens rotations 
introduce an additional zero below the diagonal of the unitary.
Thus a sequence of such rotations
realizes the inverse unitary up to relative phases,
and the reversed sequence of inverse rotations realizes the
unitary itself (up to a diagonal gate).  There are $d(d-1)/2$ 
elements below the diagonal; hence the gate count in Eq. \eqref{eq:QR}. 
The entire synthesis then follows by 
$U=D[\prod_{\ell=1}^{d(d-1)/2}G_{j_{\ell}k_{\ell}}]$.
Using an Euler decomposition of $SU(2)$, the diagonal gate can be can 
be built using $3(d-1)$ Givens rotations.

A second way to synthesize a unitary transformation
 is to use a spectral decomposition
\begin{equation}
U = \prod_{\ell=0}^{d-1} W_{\ell} C_{\ell} W_{\ell}^{\dagger}
\label{spec}
\end{equation}
where $W_{\ell}$ is a unitary matrix that maps the basis state $\ket{\ell}$ to the
eigenvector corresponding to the ${\ell}$th eigenvalue of $U$, and
$C_{\ell}$ is the identity matrix with its $({\ell},{\ell})$ element replaced
by the ${\ell}$th eigenvalue.  Each matrix $W_{\ell}^{\dagger}$ implements
a state-synthesis operation and can be implemented
as a product of $G_{jk}(\gamma,\phi)$.
The first major topic of this work is parallelism, both in state
synthesis and in the two unitary constructions above.

Particular physical systems exhibit symmetries that constrain and
refine the broad picture of unitary evolution presented so far
\cite{MuthukrishnanStroud:00,Knill_state}.
This work focuses on systems in which a limited number of pairs of states 
can be coupled at any given time.  The examplar system is a 
qudit encoded in the ground hyperfine state of a neutral alkali atom,
where the number of pairs that may be coupled at once is determined by
the number of lasers incident on the atoms.
Other candidate systems for quantum computation, such as 
flux based Josephson junction qudits and electronic states of trapped ions, 
may allow this type of control.  

We recall how selection rules on an atom with hyperfine electron structure
constrains the allowed Givens evolutions of the system 
\cite{selectionQR,spins}.
A pair of Raman pulses can couple states 
$\ket{F_{\downarrow},M_F}\leftrightarrow \ket{F_{\uparrow},M_F'}$.  In the 
linear Zeeman regime, a specific pair of hyperfine states can be addressed 
by choosing the appropriate frequency and polarization of the two Raman beams.
The coupling acts on the electron degree of freedom which imposes a 
selection rule $\Delta M_F=M_F-M_F'=\pm 2,\pm 1, 0$.  To demonstrate the power 
of our unitary synthesis technique, we restrict discussion to the selection 
rule $\Delta M_F=M_F-M_F'=\pm 1, 0$.
This restriction is valid when the detuning of 
each Raman laser beam from the excited state is much larger than the 
hyperfine splitting in the excited state $(\Delta\gg E_{ehf})$ 
\cite{Deutsch:98}.   There is a practical
advantage to restricting discussion to this selection rule.  Spontaneous 
emission 
during the Raman gate scales as 
$\gamma\sim\Gamma|\Omega_1\Omega_2|/\Delta^2$, where
$\Omega_{1,2}$ are the Rabi frequencies of the two Raman beams and $\Gamma$
is the spontaneous emission rate from the excited state.   Working in the limit
of large detunings reduces errors due to spontaneous scattering events. 

\begin{figure}
\begin{center}
\includegraphics[scale=0.33]{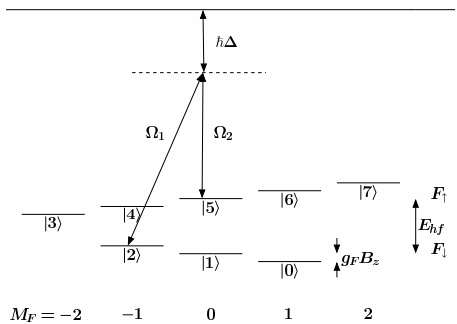}
\caption{\label{fig:hyperfine}A single $d=8$ 
qudit encoded in the ground 
state hyperfine levels of $^{87}$Rb.  A pair of lasers can couple states 
in different hyperfine manifolds according to the selection rule 
$\Delta M_F=0,\pm1$.}
\end{center}
\end{figure}

The hyperfine 
levels for a $d=8$ qudit and the induced coupling graph are shown in 
Figs. \ref{fig:hyperfine} and \ref{fig:rb}.  We assume 
that the amplitude and phase of the Raman beams can be controlled so that 
each Givens rotation
$G_{jk}(\gamma,\phi)$ can be generated in a single time step 
(see \cite{selectionQR}).  It is notable that while the multitude of hyperfine
levels in atomic systems provides a large state space of quantum 
information processing, these states 
are sensitive to errors.  For instance, it is possible to choose disjoint 
two-dimensional subspaces,
spanned by $\{\ket{F_{\downarrow},M_F},\ket{F_{\uparrow},-M_F}\}$, 
that are insensitive to small magnetic
field fluctuations along the quantization axis.  
Fluctuating fields along different axes have negligible
effect provided a large enough fixed Zeeman field is applied. 
There are no such error 
avoidance codes when using the entire hyperfine.  Hence parallelism, on 
a scale that can 
support error correction on a time scale fast compared to 
environmental noise, will be crucial.

\subsection{Achieving parallelism in state synthesis}

To implement the unitary state synthesis operator $W_{\ell}^{\dagger}$, we 
construct a sequence of rotations taking a particular
vector to a given state $\ket{\ell}$.  Again, this is technically the
reverse of state synthesis: $\;$ $W \ket{\ell}=\ket{\psi}$ for a generic
pure state $\ket{\psi}$ inverts to a sequence of unitaries
$G_{jk}(\gamma,\phi)$ accomplishing $W^\dagger \ket{\psi}=\ket{\ell}$.
Thus in the application $\ket{\ell}$ will be the fiducial state,
and we attempt to treat all possibilities.
We abbreviate the rotation of Eq.~(\ref{eq:givens}) by $G_{jk}$.

One tool for identifying sequences of rotations
that produce $W_{\ell}^{\dagger}$
is the rotation or coupling graph, in which node $j$ is connected to
node $k$ if a rotation between rows $j$ and $k$ is
physically realizable \cite{qrrestricted}.
Then $W_{\ell}^{\dagger}$ is constructed
by the sequence of rotations determined
by constructing a spanning tree rooted at $\ell$
and successively eliminating leaf nodes by a rotation
with their parent.  Recall, a spanning tree of a 
graph $G(V,E)$ connects all $d=|V|$ nodes of $G$ with exactly $d-1$ edges
from the set $E$.

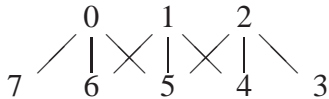
\begin{figure}
\centerline{
\begin{picture}(250,60)
\put(80,50)  {\large \bf $0$}
\put(130,50) {\large \bf $1$}
\put(180,50) {\large \bf $2$}
\put(30,5)   {\large \bf $7$}
\put(80,5)   {\large \bf $6$}
\put(130,5)  {\large \bf $5$}
\put(180,5)  {\large \bf $4$}
\put(230,5)  {\large \bf $3$}
\put( 85,20){\line(0,1){23}}  
\put(135,20){\line(0,1){23}}  
\put(185,20){\line(0,1){23}}  
\put( 75,45){{\line(-1,-1){25}}} 
\put( 95,45){{\line( 1,-1){25}}} 
\put(125,45){{\line(-1,-1){25}}} 
\put(145,45){{\line( 1,-1){25}}} 
\put(175,45){{\line(-1,-1){25}}} 
\put(195,45){{\line( 1,-1){25}}} 
\end{picture}
}
\caption{\label{fig:rb}Coupling graph for $^{87}${\tt Rb}.}
\end{figure}

Consider, for example, the coupling graph of Figure \ref{fig:rb}. To
perform state synthesis for $\ket{0}$, we can form a spanning
tree by breaking the edge between 1 and 5,
breaking one of the edges in the cycle 
$0,5,2,4,1,6,0$, and choosing the
root to be $\ket{0}$. 
If we break the edge between 2 and 4, then the resulting
tree has three leaves, 7 (eliminated by $G_{07}$), 
3 (eliminated by $G_{23}$).  
and 4 (eliminated by $G_{14}$), 
We can then eliminate the
two resulting leaves 1 and 2, and then 6 and 5.  Therefore,
we have constructed a rotation sequence
\[
G_{05} G_{06} G_{61} G_{52} G_{14} G_{23} G_{07}
\]
that synthesizes $\ket{0}$ in 7 steps.

To understand the potential for parallelism, note that
some of these rotations commute and can therefore be applied
in parallel.  This is a special case of
the assertion that infinitesimal unitaries
$ih_1,ih_2 \in \mathfrak{u}(d)$ may be applied in parallel
iff $[h_1,h_2]=0$ iff $e^{it h_1}$ and $e^{it h_2}$
commute for all $t$ real.
We rely on the following result.

\begin{proposition}
A subsequence of $p$ rotations can be applied in parallel
if and only if all $2 p$ indices are distinct.
\label{propone}
\end{proposition}

\begin{proof}
It is easy to verify that if all four indices are
distinct, then $G_{jk} G_{nm} = G_{nm} G_{jk}$. 
Conversely, if the four indices are not distinct, then the order of
application matters and therefore the rotations cannot be applied
in parallel.
The result follows by induction on $p$.
\end{proof}

Using square brackets to group rotations that can be applied in parallel,
the 7-step rotation sequence of our example becomes
the 4-step parallel rotation sequence
\begin{equation}
G_{05} G_{06} [G_{61} G_{52}] [G_{14} G_{23} G_{07}] \, .
\label{eq:zerosyn}
\end{equation}
 
\begin{figure}
\begin{center}
\includegraphics[scale=0.6]{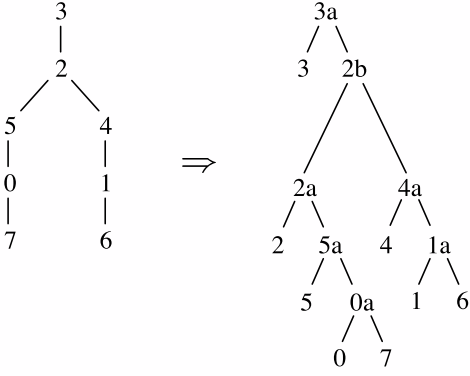}
\caption{\label{fig:bintree}A spanning tree (left) and a 
BCT (right) for node 3 of $^{87}${\tt Rb}.}
\end{center}
\end{figure}

The next interesting question is how we might determine an 
ordering of rotations to produce a parallel
rotation sequence with a small number of steps.
To answer this question,
we build upon an algorithm of He and Yesha \cite[Sec. 3.1]{heyesha}.
Given a spanning tree,
they create a {\em binary computation tree} (BCT) by working from the
bottom up and replacing every
internal node in the spanning tree by a leaf connected to a chain
of $p$ nodes, where $p$ is the number of children of the node.
They then attach one child to each of the new nodes.
The final result is a binary tree.
(This process is illustrated in Figure \ref{fig:bintree}
for a spanning tree of the coupling graph in
Figure \ref{fig:rb} rooted at node 3.)
The following proposition shows that the 
number of steps in our parallel rotation sequence
is equal to the height of the BCT,
not the height of the spanning tree.

\begin{proposition}
\label{lm:opt}
An ordering of the rotations can be obtained by
constructing the BCT for a spanning tree of the coupling graph
and scheduling
each rotation at time step $k-j$, where $k$ is the height of the
BCT and $j$ is the distance of the two leaves of the rotation from the root
of the BCT.
The resulting number of steps is $k-1$.
\end{proposition}

\begin{proof}
{In constructing the BCT, we have split each node of the spanning tree
that is involved in more than one rotation into a chain of nodes,
each on a distinct level.  This assures that rotations on the same
level commute and therefore can be applied in parallel.
}
\end{proof}

The resulting ordering is within a factor of $O(\log_2 m)$ of optimal,
where $m$ is the number of rotations \cite{heyesha}.
We next present a direct (in fact greedy) algorithm which also
orders the rotations for optimal parallelism.

At each step, consider each leaf of the spanning tree in order of
its distance from the root (more distant leaves first),
and process (remove) any leaf whose rotation
can be applied in parallel with those already chosen for processing.
The two algorithms give the same number of steps but
perhaps assign a different timing to some rotations.
For example, the greedy algorithm applied to 
the spanning tree on the left of Figure \ref{fig:bintree}
yields
\[
G_{32}G_{24}G_{25}[G_{50}G_{41}][G_{07}G_{16}] \, ,
\]
while the BCT on the right of the figure yields the
schedule
\[
G_{32} G_{24} [G_{25}G_{41}] [G_{50} G_{16}] G_{07}.
\]
Both rotation sequences require 5 steps.

Therefore, we can determine an ordering for the
rotations to perform state synthesis for $\ket{\ell}$
by considering in turn each possible
spanning tree rooted at $\ket{\ell}$, constructing an ordering
for it, and choosing the ordering that provides the smallest number of steps.

It is possible that resource constraints prevent us from
implementing a parallel ordering.  
Suppose for example a limited number of laser beams
allows us to apply only two rotations at at time.
State synthesis for $\ket{0}$
(Eq.~\ref{eq:zerosyn}) can still be accomplished using
a 4-step rotation sequence, but it requires a nontrivial
rearrangement:
\begin{equation}
G_{05} [G_{06} G_{52}] [G_{61} G_{07}] [G_{14} G_{23} ] \, .
\end{equation}
In general, such a constrained scheduling problem is difficult to solve
exactly, although good heuristics exist.

\subsection{Examples of parallelism in state synthesis}

We apply our state synthesis algorithms to rubidium and cesium.

\paragraph{Hyperfine levels of $^{87}${\tt Rb}.}

Only the 9 transitions corresponding
to the edges of the coupling graph of Figure \ref{fig:rb} 
are allowed, and the edge between 1 and 5 will
not be used in our algorithms, since it does not lead to 
speed-up \cite{footnote1}.

Optimal parallel rotation sequences, constructed using
Proposition \ref{lm:opt}, are given in Table \ref{tab:rb}.
They require 5 steps for $\ket{3}$ and $\ket{7}$ and
4 steps for the other kets, rather than the 7 steps of the
sequential algorithm.

\begin{table}{}
\caption{\label{tab:rb}Parallel rotation sequences for state synthesis
using laser Raman coupled connections between 
hyperfine states of $^{87}${\tt Rb}.}
\begin{centering}
\begin{tabular}{l  r }
$\ket{0}$&$G_{05}$ $[G_{06}G_{52}]$ $[G_{61}G_{07}]$ $[G_{14}G_{23}]$ \\
$\ket{1}$&$G_{16}$ $[G_{60}G_{14}]$ $[G_{05}G_{42}]$ $[G_{23}G_{07}]$ \\
$\ket{2}$&$G_{25}$ $[G_{24}G_{50}]$ $[G_{41}G_{23}]$ $[G_{16}G_{07}]$ \\
$\ket{3}$&$G_{32}$ $G_{24}$ $G_{25}$ $[G_{50}G_{41}]$ $[G_{16}G_{07}]$ \\
$\ket{4}$&$G_{41}$ $[G_{16}G_{42}]$ $[G_{25}G_{60}]$ $[G_{23}G_{07}]$ \\
$\ket{5}$&$G_{50}$ $[G_{52}G_{07}]$ $[G_{06}G_{24}]$ $[G_{61}G_{23}]$ \\
$\ket{6}$&$G_{61}$ $[G_{14}G_{60}]$ $[G_{05}G_{42}]$ $[G_{23}G_{07}]$ \\
$\ket{7}$&$G_{70}$ $G_{06}$ $[G_{61}G_{05}]$ $[G_{14}G_{52}]$ $G_{23}$ \\
\end{tabular}\\
\end{centering}
\end{table}

\paragraph{Hyperfine levels of $^{133}${\tt Cs}.}

\begin{figure}
\centerline{
\begin{picture}(475,60)
\put(80,50)  {\large \bf $0$}
\put(130,50) {\large \bf $1$}
\put(180,50) {\large \bf $2$}
\put(230,50) {\large \bf $3$}
\put(280,50) {\large \bf $4$}
\put(330,50) {\large \bf $5$}
\put(380,50) {\large \bf $6$}
\put(30,5)   {\large \bf $15$}
\put(80,5)   {\large \bf $14$}
\put(130,5)  {\large \bf $13$}
\put(180,5)  {\large \bf $12$}
\put(230,5)  {\large \bf $11$}
\put(280,5)  {\large \bf $10$}
\put(330,5)  {\large \bf $9$}
\put(380,5)  {\large \bf $8$}
\put(430,5)  {\large \bf $7$}
\put( 85,20){\line(0,1){23}}  
\put(135,20){\line(0,1){23}}  
\put(185,20){\line(0,1){23}}  
\put(235,20){\line(0,1){23}}  
\put(285,20){\line(0,1){23}}  
\put(335,20){\line(0,1){23}}  
\put(385,20){\line(0,1){23}}  
\put( 75,45){\cred{\line(-1,-1){25}}} 
\put( 95,45){\cred{\line( 1,-1){25}}} 
\put(125,45){\cblu{\line(-1,-1){25}}} 
\put(145,45){\cblu{\line( 1,-1){25}}} 
\put(175,45){\cred{\line(-1,-1){25}}} 
\put(195,45){\cred{\line( 1,-1){25}}} 
\put(225,45){\cblu{\line(-1,-1){25}}} 
\put(245,45){\cblu{\line( 1,-1){25}}} 
\put(275,45){\cred{\line(-1,-1){25}}} 
\put(295,45){\cred{\line( 1,-1){25}}} 
\put(325,45){\cblu{\line(-1,-1){25}}} 
\put(345,45){\cblu{\line( 1,-1){25}}} 
\put(375,45){\cred{\line(-1,-1){25}}} 
\put(395,45){\cred{\line( 1,-1){25}}} 
\end{picture}
}
\caption{\label{fig:cestrans}Coupling graph for $^{133}${\tt Cs}.}
\end{figure}
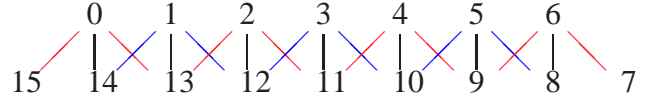

The coupling graph of allowed transitions for $^{133}${\tt Cs}
is given in Figure \ref{fig:cestrans}.  We partition
these transitions into three groups:
\bl
\ii The {\em outer chain} of (red) transitions between
$\ket{15}$, $\ket{0}$, $\ket{13}$, $\ket{2}$, $\ket{11}$, $\ket{4}$, 
$\ket{9}$, $\ket{6}$, and $\ket{7}$.
\ii The {\em inner chain} of (blue) transitions between
$\ket{14}$, $\ket{1}$, $\ket{12}$, $\ket{3}$, $\ket{10}$, $\ket{5}$, 
and $\ket{8}$.
\ii A {\em ladder} of transitions between the two chains.
\el

Since $d=16$, 
state synthesis requires 15 rotations.  If the desired
state is $\ket{3}$, for example, then we can use the
outer chain of transitions to depopulate
$\ket{7}$, $\ket{6}$, $\ket{9}$, $\ket{4}$ (in order)
and then $\ket{15}$, $\ket{0}$, $\ket{13}$, $\ket{2}$,
and then use the ladder transition from $\ket{11}$
to $\ket{3}$.  Similarly, the inner chain of transitions
can be used to empty
$\ket{14}$, $\ket{1}$, $\ket{12}$, $\ket{8}$, $\ket{5}$,
and finally $\ket{10}$.
This pattern of using the outer chain, the inner chain, 
and a single ladder transition accomplishes state
synthesis for an arbitrary state.

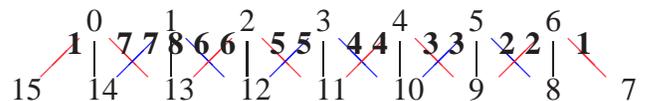
\begin{figure}[b]
\centerline{
\begin{picture}(475,60)
\put(80,50)  {\large \bf $0$}
\put(130,50) {\large \bf $1$}
\put(180,50) {\large \bf $2$}
\put(230,50) {\large \bf $3$}
\put(280,50) {\large \bf $4$}
\put(330,50) {\large \bf $5$}
\put(380,50) {\large \bf $6$}
\put(30,5)   {\large \bf $15$}
\put(80,5)   {\large \bf $14$}
\put(130,5)  {\large \bf $13$}
\put(180,5)  {\large \bf $12$}
\put(230,5)  {\large \bf $11$}
\put(280,5)  {\large \bf $10$}
\put(330,5)  {\large \bf $9$}
\put(380,5)  {\large \bf $8$}
\put(430,5)  {\large \bf $7$}
\put( 85,20){\line(0,1){23}}  
\put(135,43){\line(0,-1){23}}  
\put(133,35){\large \bf 8}
\put(185,20){\line(0,1){23}}  
\put(235,20){\line(0,1){23}}  
\put(285,20){\line(0,1){23}}  
\put(335,20){\line(0,1){23}}  
\put(385,20){\line(0,1){23}}  
\put( 75,45){\cred{\line(-1,-1){25}}} 
\put( 67,35){\large \bf 1}
\put( 95,45){\cred{\line( 1,-1){25}}} 
\put(100,35){\large \bf 7}
\put(125,45){\cblu{\line(-1,-1){25}}} 
\put(117,35){\large \bf 7}
\put(145,45){\cblu{\line( 1,-1){25}}} 
\put(150,35){\large \bf 6}
\put(175,45){\cred{\line(-1,-1){25}}} 
\put(167,35){\large \bf 6}
\put(195,45){\cred{\line( 1,-1){25}}} 
\put(200,35){\large \bf 5}
\put(225,45){\cblu{\line(-1,-1){25}}} 
\put(217,35){\large \bf 5}
\put(245,45){\cblu{\line( 1,-1){25}}} 
\put(250,35){\large \bf 4}
\put(275,45){\cred{\line(-1,-1){25}}} 
\put(267,35){\large \bf 4}
\put(295,45){\cred{\line( 1,-1){25}}} 
\put(300,35){\large \bf 3}
\put(325,45){\cblu{\line(-1,-1){25}}} 
\put(317,35){\large \bf 3}
\put(345,45){\cblu{\line( 1,-1){25}}} 
\put(350,35){\large \bf 2}
\put(375,45){\cred{\line(-1,-1){25}}} 
\put(367,35){\large \bf 2}
\put(395,45){\cred{\line( 1,-1){25}}} 
\put(400,35){\large \bf 1}
\end{picture}
}
\caption{\label{fig:st2a}State synthesis for $\ket{1}$ for the Cesium alkali
using two-way parallelism. All transitions
are directed toward $\ket{1}$.}
\end{figure}

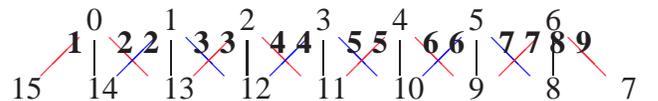
\begin{figure}[b]
\centerline{
\begin{picture}(475,60)
\put(80,50)  {\large \bf $0$}
\put(130,50) {\large \bf $1$}
\put(180,50) {\large \bf $2$}
\put(230,50) {\large \bf $3$}
\put(280,50) {\large \bf $4$}
\put(330,50) {\large \bf $5$}
\put(380,50) {\large \bf $6$}
\put(30,5)   {\large \bf $15$}
\put(80,5)   {\large \bf $14$}
\put(130,5)  {\large \bf $13$}
\put(180,5)  {\large \bf $12$}
\put(230,5)  {\large \bf $11$}
\put(280,5)  {\large \bf $10$}
\put(330,5)  {\large \bf $9$}
\put(380,5)  {\large \bf $8$}
\put(430,5)  {\large \bf $7$}
\put( 85,20){\line(0,1){23}}  
\put(135,43){\line(0,-1){23}}  
\put(185,20){\line(0,1){23}}  
\put(235,20){\line(0,1){23}}  
\put(285,20){\line(0,1){23}}  
\put(335,20){\line(0,1){23}}  
\put(385,20){\line(0,1){23}}  
\put(383,35){\large \bf 8}
\put( 75,45){\cred{\line(-1,-1){25}}} 
\put( 67,35){\large \bf 1}
\put( 95,45){\cred{\line( 1,-1){25}}} 
\put(100,35){\large \bf 2}
\put(125,45){\cblu{\line(-1,-1){25}}} 
\put(117,35){\large \bf 2}
\put(145,45){\cblu{\line( 1,-1){25}}} 
\put(150,35){\large \bf 3}
\put(175,45){\cred{\line(-1,-1){25}}} 
\put(167,35){\large \bf 3}
\put(195,45){\cred{\line( 1,-1){25}}} 
\put(200,35){\large \bf 4}
\put(225,45){\cblu{\line(-1,-1){25}}} 
\put(217,35){\large \bf 4}
\put(245,45){\cblu{\line( 1,-1){25}}} 
\put(250,35){\large \bf 5}
\put(275,45){\cred{\line(-1,-1){25}}} 
\put(267,35){\large \bf 5}
\put(295,45){\cred{\line( 1,-1){25}}} 
\put(300,35){\large \bf 6}
\put(325,45){\cblu{\line(-1,-1){25}}} 
\put(317,35){\large \bf 6}
\put(345,45){\cblu{\line( 1,-1){25}}} 
\put(350,35){\large \bf 7}
\put(375,45){\cred{\line(-1,-1){25}}} 
\put(367,35){\large \bf 7}
\put(395,45){\cred{\line( 1,-1){25}}} 
\put(400,35){\large \bf 9}
\end{picture}
}
\caption{\label{fig:st2b}State synthesis for $\ket{7}$ for the Cesium alkali
using two-way parallelism.}
\end{figure}

Complete parallelism is possible in the application of
rotations from the outer chain with those in the inner,
since no state is involved in both chains.
If two rotations can be applied at once, then
we need 9 steps for state synthesis to $\ket{15}$
or $\ket{7}$ and 8 steps for the other kets.
We illustrate such a scheme in Figures \ref{fig:st2a}
and \ref{fig:st2b}, marking each transition with the step
at which it is used.

\subsection{Parallelism in one-qudit unitary processes}
\label{uptodiag}
Recall that a state synthesis routine yields routines for
realizing arbitrary one-qudit unitary evolutions in (at least)
two different ways: $\;$ by invoking the $QR$ matrix decomposition
(Eq.~\ref{eq:QR}) or by the spectral theorem (Eq.~\ref{spec}).
The number of parallel steps for a generic unitary
can be significantly greater when using the spectral theorem.
For example, for $^{87}${\tt Rb}, the
spectral decomposition
would take 68 steps plus the steps needed to apply the phases.
The number of steps to apply parallel $QR$ is much less; with 3-way parallelism
it is  at most $2n-3=13$ ($n=8$) plus the steps to apply the phases.  
Also note that the sequential $QR$ requires $n(n-1)/2=28$ steps, so this
is a considerable speedup.

A rotation sequence that achieves this bound of 13 steps for $QR$
can be constructed using the precedence graph for the
computation \cite{olst}.
Suppose we order the rows as $7, 5, 0, 6, 1, 4, 2, 3$. 
We usually use rotations that eliminate an element in any row
by a rotation with the element directly above it, but in the first
column we use the rotation sequence
\[
G_{70} G_{05} G_{06} [G_{52} G_{61}] [G_{14} G_{23} ].
\]
This sequence specifies predecessors for each rotation
in the first column.
Define the predecessors of a rotation for columns after the first
to be the rotations zeroing elements to the
south, west and northwest, if those rotations exist.
Each rotation can be performed after all of its predecessors are
completed.
Therefore, the numerical value of each entry 
below the diagonal
in the following matrix denotes the step at which the entry can be zeroed:

{\small 
\[
\begin{array}{c} 7\\5\\0\\6\\1\\4\\2\\3 \end{array}
\bmx{r r r r r r r r}
 x &  x &  x&  x &  x &  x &  x &  x \\
 4 &  x &  x&  x &  x &  x &  x &  x \\
 5 &  8 &  x&  x &  x &  x &  x &  x \\
 3 &  7 &  9&  x &  x &  x &  x &  x \\
 2 &  6 &  8& 10 &  x &  x &  x &  x \\
 1 &  5 &  7&  9 & 11 &  x &  x &  x \\
 2 &  4 &  6&  8 & 10 & 12 &  x &  x \\
 1 &  3 &  5&  7 &  9 & 11 & 13 &  x \\
\emx \, .
\]
}

\noindent
Thus, using 3-way parallelism,
an arbitrary unitary can be applied in 13 steps, plus the
steps for phasing.

If only 2-way parallelism is 
allowed, then more steps are necessary.
We schedule rotations by cycling through
the columns in round-robin order (right to left),
scheduling at most one rotation per column,
until all rotations are scheduled. 
If the predecessors of the column's next rotation are scheduled,
then that rotation is scheduled for the earliest available time
step after their scheduled steps.
The resulting time steps are:
{\small
\[
\begin{array}{c} 7\\5\\0\\6\\1\\4\\2\\3 \end{array}
\bmx{r r r r r r r r}
 x &  x &  x&  x &  x &  x &  x &  x \\
 4 &  x &  x&  x &  x &  x &  x &  x \\
 6 & 10 &  x&  x &  x &  x &  x &  x \\
 3 &  8 & 11&  x &  x &  x &  x &  x \\
 2 &  7 &  9& 12 &  x &  x &  x &  x \\
 1 &  5 &  8& 11 & 13 &  x &  x &  x \\
 2 &  4 &  6&  9 & 12 & 14 &  x &  x \\
 1 &  3 &  5&  7 & 10 & 13 & 15 &  x \\
\emx \, .
\]
}

\noindent
These $15$ steps are optimal for $2$-way parallelism;
the last two rotations
must be applied sequentially, so the $28$ rotations cannot be applied in
$14$ steps.

A similar construction using the Cesium coupling graph shows that 
at most 29 steps are required using 7-way parallelism.
We order the rows as 15,14,0,13,1,12,2,11,3,10,4,9,5,8,6,7.
The rotations used in the first column are
\[
\begin{array}{c}
G_{15,0} G_{0,13} G_{13,2} G_{2,11} G_{11,4} G_{4,9} G_{9,5} G_{9,6} \\
$$ [G_{0,14} G_{13,1} G_{2,12} G_{11,3} G_{4,10} G_{5,8} G_{6,7}],
\end{array}
\]
while in other columns
we use rotations that eliminate an element in any row
by a rotation with the element directly above it. 
The time steps are as follows:

{\footnotesize 
\[
\begin{array}{r} 15\\14\\0\\13\\1\\12\\2\\11\\3\\10\\4\\9\\5\\8\\6\\7
                                        \end{array}
\bmx{r r r r r r r r r r r r r r r r}
x &  x &  x&  x &  x &  x &  x &  x  &  x &  x &  x&  x &  x &  x &  x &  x \\
 1 &  x &  x&  x &  x &  x &  x &  x  &  x &  x &  x&  x &  x &  x &  x &  x \\
 9 & 16 &  x&  x &  x &  x &  x &  x  &  x &  x &  x&  x &  x &  x &  x &  x \\
 8 & 15 & 17&  x &  x &  x &  x &  x  &  x &  x &  x&  x &  x &  x &  x &  x \\
 1 & 14 & 16& 18 &  x &  x &  x &  x  &  x &  x &  x&  x &  x &  x &  x &  x \\
 1 & 13 & 15& 17 & 19 &  x &  x &  x  &  x &  x &  x&  x &  x &  x &  x &  x \\
 7 & 12 & 14& 16 & 18 & 20 &  x &  x  &  x &  x &  x&  x &  x &  x &  x &  x \\
 6 & 11 & 13& 15 & 17 & 19 & 21 &  x  &  x &  x &  x&  x &  x &  x &  x &  x \\
 1 & 10 & 12& 14 & 16 & 18 & 20 & 22  &  x &  x &  x&  x &  x &  x &  x &  x \\
 1 &  9 & 11& 13 & 15 & 17 & 19 & 21  & 23 &  x &  x&  x &  x &  x &  x &  x \\
 5 &  8 & 10& 12 & 14 & 16 & 18 & 20  & 22 & 24 &  x&  x &  x &  x &  x &  x \\
 4 &  7 &  9& 11 & 13 & 15 & 17 & 19  & 21 & 23 & 25&  x &  x &  x &  x &  x \\
 3 &  6 &  8& 10 & 12 & 14 & 16 & 18  & 20 & 22 & 24& 26 &  x &  x &  x &  x \\
 1 &  5 &  7&  9 & 11 & 13 & 15 & 17  & 19 & 21 & 23& 25 & 27 &  x &  x &  x \\
 2 &  4 &  6&  8 & 10 & 12 & 14 & 16  & 18 & 20 & 22& 24 & 26 & 28 &  x &  x \\
 1 &  3 &  5&  7 &  9 & 11 & 13 & 15  & 17 & 19 & 21& 23 & 25 & 27 & 29 &  x \\
\emx
\]
}

If fewer parallel resources are available, we can again reschedule our
steps as done above for Rubidium.
For 3-way parallelism, for example, we can schedule the rotations as

{\footnotesize 
\[
\begin{array}{r} 15\\14\\0\\13\\1\\12\\2\\11\\3\\10\\4\\9\\5\\8\\6\\7
                                        \end{array}
\bmx{r r r r r r r r r r r r r r r r}
x &  x &  x&  x &  x &  x &  x &  x  &  x &  x &  x&  x &  x &  x &  x &  x \\
 4 &  x &  x&  x &  x &  x &  x &  x  &  x &  x &  x&  x &  x &  x &  x &  x \\
19 & 24 &  x&  x &  x &  x &  x &  x  &  x &  x &  x&  x &  x &  x &  x &  x \\
17 & 22 & 26&  x &  x &  x &  x &  x  &  x &  x &  x&  x &  x &  x &  x &  x \\
 3 & 19 & 24& 28 &  x &  x &  x &  x  &  x &  x &  x&  x &  x &  x &  x &  x \\
 2 & 17 & 21& 26 & 30 &  x &  x &  x  &  x &  x &  x&  x &  x &  x &  x &  x \\
11 & 15 & 19& 23 & 28 & 32 &  x &  x  &  x &  x &  x&  x &  x &  x &  x &  x \\
 6 & 13 & 16& 21 & 26 & 30 & 34 &  x  &  x &  x &  x&  x &  x &  x &  x &  x \\
 2 & 11 & 14& 18 & 23 & 28 & 32 & 35  &  x &  x &  x&  x &  x &  x &  x &  x \\
 1 & 10 & 13& 16 & 21 & 25 & 30 & 33  & 36 &  x &  x&  x &  x &  x &  x &  x \\
 5 &  8 & 11& 14 & 18 & 23 & 27 & 31  & 35 & 37 &  x&  x &  x &  x &  x &  x \\
 4 &  7 &  9& 12 & 16 & 20 & 25 & 29  & 33 & 36 & 38&  x &  x &  x &  x &  x \\
 3 &  6 &  8& 10 & 14 & 18 & 22 & 27  & 31 & 34 & 37& 39 &  x &  x &  x &  x \\
 1 &  5 &  7&  9 & 12 & 15 & 20 & 25  & 29 & 33 & 36& 38 & 40 &  x &  x &  x \\
 2 &  4 &  6&  8 & 10 & 13 & 17 & 22  & 27 & 31 & 34& 37 & 39 & 41 &  x &  x \\
 1 &  3 &  5&  7 &  9 & 12 & 15 & 20  & 24 & 29 & 32& 35 & 38 & 40 & 42 &  x \\
\emx
\]
}

\begin{table}
\caption{\label{tab:2}  Number of parallel steps to synthesize
a generic unitary operation $U$, up to a diagonal gate $D$,
on a single atomic qudit. 
Each Raman pair of laser beams counts as a single resource and the logical 
depth is the number of parallel Raman gate sequences needed in our
 $QR$ diagonalization of $U$. 
The number in parenthesis is our best lower bound.
The tradeoff between time and resources is evident.}
\begin{ruledtabular}
\begin{tabular}{ccc}
\hline
Parallelism  & Logical Depth:   $^{87}$Rb (d=8) & $^{133}$Cs (d=16)  \\
\hline
7-way   &\quad\quad\quad\quad  $13$ (11) & $29$ (26) \\
\hline
6-way   &\quad\quad\quad\quad  $13$ (11) & $30$ (26) \\
\hline
5-way   &\quad\quad\quad\quad  $13$ (11) & $31$ (26) \\
\hline
4-way   &\quad\quad\quad\quad  $13$ (11) & $35$ (26) \\
\hline
3-way   &\quad\quad\quad\quad  $13$ (11) & $42$ (42) \\
\hline
2-way   &\quad\quad\quad\quad  $15$ (15) & $62$ (61) \\
\hline
1-way   &\quad\quad\quad\quad  $28$ (28) & $120$ (120) \\
\hline
\end{tabular}
\end{ruledtabular}
\end{table}

A summary of the tradeoff between resources and gate times 
with qudits encoded in ground hyperfine levels of $^{87}$Rb and $^{133}$Cs
 is given in Table \ref{tab:2}. 
As noted above, the 2-way construction of 15 steps for $^{87}$Rb is optimal.
Similar reasoning gives the 2- and 3-way lower bounds for $^{133}$Cs;
for example, 118 rotations divided by 3 gives 40 steps plus two final steps
for the last two rotations.
The other lower bounds in the table
are obtained assuming a completely connected 
coupling graph and  ($n/2$)-way parallelism.  In that case, 
if $n=2^p$, we
can insert $n/2$ zeros in the first column at step 1, 
up to $n/4$ zeros in the first two columns 2 at step 2, ... , 
$1$ zero in the first $p$ columns  at step $p$, and then start 
 the reduction in the $j$th column for $j=p+1,\dots,n-1$ 
 at step $\log_2 n + 2(j-\log_2 n)$, for a total of 
$\log_2 n + 2(n-1-\log_2 n)$
steps.
Other choices of rotation
sequences may reduce some entries in the table.

\subsection{Parallel diagonal gates}

Up to this point our discussion has counted the number of parallel steps 
needed to construct any single-qudit unitary up to a diagonal gate $D$.  
Synthesizing the diagonal gate is unnecessary if the target qudit will 
remain dormant until a measurement in the computational basis.  
However, if the 
qudit will be targetted by subsequent operations 
then it will be necessary
to phase the basis states of the qudit appropriately. 
We next consider parallel constructions for $D$.  There are two
variations of this problem to discuss.  In the first, we 
define a gate to be an evolution
by the generator $\lambda_{j,k}^z$,
where $j$ and $k$ are paired levels.
In the second, the gate 
library is restricted to Givens rotations (Eq. \ref{eq:givens}) 
as is the case in systems
controlled with Raman laser pairs. Here one cannot realize a 
diagonal Hamiltonian
directly but rather may simulate $\mbox{e}^{it \lambda^z_{j,k}}$
using an Euler angle decomposition.

First, note that the $D$ gate itself need only be simulated 
up to a local phase: e.g., we may chose $D\in SU(d)$.  
Simulating a diagonal gate with $d-1$ independent phases should require
appropriate couplings between $d-1$ pairs of states.  There is a 
large amount of freedom in
the choice of the set of the $d-1$ state pairs:  
any $D\in SU(d)$ can be 
written $D=\prod_{m=1}^{d-1}e^{i\phi_{j_m, k_m}\lambda^z_{j_m, k_m}}$, 
provided the set of edges $E = \{(j_m,k_m)\}$ creates 
a spanning tree of the coupling graph.  For
$\{ i \lambda^z_{j,k} \; : \; (j,k)\in E \}$ spans the diagonal
subalgebra of $\mathfrak{su}(d)$, and therefore we may
construct $\{\phi_{j_m,k_m}\}$ 
by solving $d-1$ linear equations \cite{selectionQR}.   
Since diagonal gates commute, the simulation (in terms of $\lambda_{j,k}^z$)
is maximally parallel, requiring one step.  
If only $k$-wise parallelism is allowed, then the number of steps is  
$\lceil (n-1)/k\rceil$.  

We next consider the case that only $\lambda^x_{j,k}$ and
$\lambda^y_{j,k}$ are allowed.  
Again choose any spanning tree for the coupling graph 
and construct $\{\phi_{j_m,k_m}\}$ by solving $d-1$ linear equations.
Color the edges of the tree so that no node has two edges
of the same color.
(For example, in Figure \ref{fig:bintree}
we need 3 colors because node 2 has 3 edges.)
Now for any edge $(j,k)$, we may indeed
realize 
$\mbox{e}^{i \phi_{j,k} t \lambda^z_{j,k}}=
\mbox{e}^{i t_1(j,k) \lambda^x_{j,k}}
\mbox{e}^{i t_2(j,k) \lambda^y_{j,k}}
\mbox{e}^{i t_3(j,k) \lambda^x_{j,k}}$ 
for appropriate timings.  Evolutions
$\mbox{e}^{i t_1 \lambda_{j,k}^x}$ and 
$\mbox{e}^{i t_2 \lambda_{j,k}^y}$ do not commute and may
not be applied in the same time step.  Yet we may group
the evolutions for a single color -- black, for example  -- 
in three time steps as
\begin{equation}
{\small
\begin{array}{l}
\big[ \prod_{(j,k) \mbox{\footnotesize \ black}} 
\mbox{e}^{i \phi_{j,k} t_1(j,k) \lambda^x_{j,k}} \big] 
\big[ \prod_{(j,k) \mbox{\footnotesize \ black}} 
\mbox{e}^{i \phi_{j,k} t_2(j,k) \lambda^y_{j,k}} \big] \\
\times \big[ \prod_{(j,k) \mbox{\footnotesize \ black}} 
\mbox{e}^{i \phi_{j,k} t_3(j,k) \lambda^x_{j,k}} \big] \, .
\end{array}
}
\end{equation}
Given a sufficient number of operations per step,
this realizes $D$ in $3c$ parallel steps, where $c$ is the
number of colors,
regardless of the number of levels in the spanning tree.
Hence, the construction is optimized by choosing a spanning 
tree that minimizes the number of colors.
The number of colors $c$ is bounded by the maximum valency $c_m$ of any node
in the coupling graph; if the coupling graph itself is a tree, then
the number of colors is exactly $c_m$.  
When control resources are limited, we make a similar
coloring, but limit the number of edges of a given
color to the maximum number of operations allowed per step.

The spanning tree of Figure \ref{fig:bintree} for $^{87}$Rb
requires three colors for the edges.
A diagonal computation can be done with the gate sequence
$D=
e^{i\phi_{3,2}\lambda^z_{3,2}}
[e^{i\phi_{2,5}\lambda^z_{2,5}}e^{i\phi_{0,7}\lambda^z_{0,7}}
 e^{i\phi_{4,1}\lambda^z_{4,1}}]
[e^{i\phi_{5,0}\lambda^z_{5,0}}e^{i\phi_{2,4}\lambda^z_{2,4}}
 e^{i\phi_{1,6}\lambda^z_{1,6}}]$, 
which requires $9$ parallel Raman pulse sequences.
Similarly, Cesium requires $9$ parallel Raman pulse
sequences.

The above treatment works for synthesizing an 
arbitrary diagonal gate $D$ without prior processing.
However, generically, the gate $D$ follows the diagonalization process 
process described in Sec. \ref{uptodiag}.  In that case 
some of pairwise
phasing operations can be subsumed in earlier steps therefore reducing
the total number of Raman pulse sequences.  First, since
Proposition \ref{propone} can be extended to any unitary, not just 
rotations of the form $G_{jk}$,
we can apply a phase correction using edge $(j,k)$ as soon as we are
finished with those two rows in the diagonalization.
Second, we are allowed to choose an edge set
for phasing different than the one we used for diagonalization.
For example, using 3-way parallelism for Rubidium,
at times $11, 12$, and $13$ of the diagonalization,
we can apply a phase correction using edge $(0,6)$;
at times $14$, $15$, and $16$ we can use $(7,0), (6,1)$, and $(5,2)$;
and at times $17, 18$, and $19$ we can finish by using $(0,5), (1,4)$, and
$(2,3)$.
A similar idea works for Cesium using 7-way parallelism:
at times $27, 28$, and $29$ use edge $(0,14)$;
at times $30, 31$, and $32$ use
$(15,0), (14,1), (13,2), (12,3), (11,4), (10,5)$, and $(9,6)$;
and at times $33, 34$, and $35$ use
$(0,13), (1,12), (2,11), (3,10), (4,9), (5,8)$, and $(6,7)$.
In Rubidium, six
extra Raman pulse sequences is optimal for phasing when nodes $\ket{2}$ 
and $\ket{3}$ are involved in the last rotation,
and six pulses ending on $\ket{6}$ and $\ket{7}$
is optimal for Cesium.
We require no more than three and six simultaneous couplings
respectively, which is also the number required for optimal
diagonalization.

\section{Parallelized non-local two-qudit gates}
\label{remote}

In this section we propose an implementation of
an arbitrary non-local unitary $U\in U(d^2)$ between two qudits $A$ and $B$.  
We suppose the qudits are spatially separated in some quantum
computing architecture, yet this architecture has the capability
to (i) prepare a large reservoir of maximally entangled ($d=2$) qubits
and (ii) the ability to shuttle halves of such Bell pairs so that
they are spatially close to qudits $A$ and $B$.  Hence, part of the
costing is the number of such Bell pairs (e-bits) consumed in the
nonlocal gate.  To be clear, we describe only a nonlocal
two-qudit gate rather than a teleported two-qudit gate meaning 
that quantum operations are performed on two qudits rather than four.  The
optimization of such a nonlocal gate presented here arises
by considering its component rotations in terms of the $QR$ decomposition.

Before stating the protocol, we argue for why it is needed.
Two criteria must be satisfied to realize 
high performance two-qudit gates.
First, nonlocality itself is desirable; most quantum computer architectures 
impose spatial limitations on inter-qudit couplings. 
It is very inconvenient to simply accept this limitation,
since fault tolerant computation requires connectivity \cite{Oskin:02}.  
Now one might also suggest directly swapping qudits in order to
achieve the required connectivity.  Yet the swap gate itself may be 
faulty, and thus the resources required to make swapping fault tolerant 
might be prohibitive.  

Second, reliable computation requires high fidelity two-qudit gates. 
Usually, Hamiltonians capable of entangling distinct qudits
are difficult to engineer (at any fidelity) and would require effort to
optimize for fidelity.  Thus, one would likely choose a particular
physically available entangling two-qudit Hamiltonian, e.g. perhaps
the controlled-phase gate $P_0=e^{i\pi \ket{0}\bra{0}\otimes \ket{0}\bra{0}}$,
and then exploit this with local unitary similarity transforms to
achieve arbitrary Givens rotations between qudit levels.
The entire process might simulate any $U\in U(d^2)$ \cite{selectionQR}.
Local unitary similarity transforms arose naturally in this
discussion, and it further implies that two-qudit nonlocality in such a scheme
would follow, given a nonlocal protocol for a single
entangling Hamiltonian.

It is difficult to design an architecture for two-qudit unitaries
which allows for both high-fidelity and high connectivity.  Some possibilities
 are noteworthy. 
As opposed to a chain of swapping operations, distant qudits might 
be swapped using
entanglement resources.  Then a non-local gate between qudits $A$ 
and $B$ can be 
done by teleporting $A$ to a location neighboring $B$, performing an 
entangling gate between $A$ and $B$ and teleporting back.  
Typically, entangled qudits (e-dits) rather than e-bits are used to teleport
qudits; i.e. each teleportation is performed with the assistance of a
maximally entangled two-qudit resource
$\ket{\Phi^+_d}=\frac{1}{\sqrt{d}}\sum_{j=0}^{d-1}\ket{j}\ket{j}$
\cite{Bennett:93}.  
While the amount of entanglement consumed using the resource $\ket{\Phi^+_d}$ 
 is low, i.e.
one e-dit$=\log(d)$ e-bits, such a protocol would still 
require high fidelity (local)
two-qudit gates between $A$ and $B$.  As hinted at
in the first paragraph of this section, a second
alternative is to teleport the gate itself using an adaptation 
of the two-qubit gate teleportation protocol \cite{Gottesman, Eisert}
\cite[\S2]{jozsa}.  In such an implementation one would build
a generic two
qudit gate between $A$ and $B$ using multiple applications of a gate 
teleport sequence
where each sequence consumed two e-dits.  Such a protocol would require 
the preparation 
of high fidelity e-dits and the implementation of generalized two-qudit 
Bell-measurements between
a memory qudit and one half of an e-dit.  

Here we describe a simple protocol for implementing a nonlocal two qudit 
gate, which has the advantage that one
need only prepare high fidelity e-bits.   
If several qubits can be controlled 
together, the entire non-local gate can be parallelized to reduce the overall 
implementation time by a factor of $O(d)$.

\subsection{A non-local controlled unitary gate}

Consider a one-qudit unitary gate controlled on dit $(d-1)$: 
\[
\wedge_1(V)\ = \ \sum_{j=0}^{d-2}\ket{j} 
\bra{j}\otimes {\bf 1}_{d}+\ket{d-1}\bra{d-1}\otimes V \, .
\] 
We label the control qudit $A$ and the target qudit $B$.
This subsection describes how such a 
gate can be implemented using 
\begin{enumerate}
\item operators local to $A$ and $B$ 
\item an e-bit.  The ancilliary e-bit is encoded in a pair of qubits, 
say $A_1$ and $B_1$, again with $A_1$ neighboring $A$ and 
$B_1$ neighboring $B$.  The joint state of the ancilla is the Bell pair
$\ket{\Phi^+}=(1/\sqrt{2})\;(\ket{00}+\ket{11})_{A_1,B_1}$.
\item 
\label{it:three}
a controlled-not gate controlled on the qudit
and targeting an ancilliary qubit.  As a formula, this gate
is $\wedge_1(\sigma^x)\ = \ \sum_{j=0}^{d-2}\ket{j} 
\bra{j}\otimes {\bf 1}_{2}-\ket{d-1}\bra{d-1}\otimes \sigma^x$.
\item 
\label{it:four}
a spatially local controlled $V$ gate with control an ancilla
bit.  As a formula, this is
(also, confusingly) 
\hbox{$\wedge_1(V)=\ket{0}\bra{0}\otimes {\bf 1}_d + 
\ket{1}\bra{1}\otimes V$}.
\end{enumerate}
The controlled gate of item \ref{it:three}
should be considered to be a primitive,
highly engineered as discussed in the previous section.  The controlled gate
of item \ref{it:four} might be decomposed into local gates
and the gate of item \ref{it:three}
using standard techniques 
\cite{MuthukrishnanStroud:00,Knill_state,selectionQR}.

The procedure for realizing $\wedge_1(V)$ is as follows.
\begin{itemize}
\item  Apply
$\wedge_1(\sigma^x_{A_1})$ with $A$ as control and $A_1$ as target.
\item  Measure $({\bf 1}_2+\sigma^z_{A_1})/2$.  
Send the one bit (c-bit) classical measurement result, $m_1$, 
to the side of qudit $B$.
\item  Perform
$e^{i\pi m_1 \sigma^x_{B_1}/2}$ on the $B$ side of the architecture.
\item  Apply the operation 
$\ket{0}\bra{0}\otimes {\bf 1}_d+\ket{1}\bra{1}\otimes V$ with 
$B_1$ as control and $B$ as target.  
\item  Measure 
$({\bf 1}_2+\sigma^x_{B_1})/2$ and send the c-bit measurement result 
$m_2$ to $A$.  
\item  Apply the a relative phase to state $d-1$ of $A$ iff
$m_2=1$, i.e. apply
$P_{d-1}=e^{i\pi m_2 \ket{d-1}_{A}{_{A}}\bra{d-1}}$. 
\end{itemize}

\subsection{Bootstrap to nonlocal two-qudit state synthesis}

We next consider the question of building a nonlocal two-qudit
state synthesis operator.  We may write any two-qudit state
$\ket{\psi}=\sum_{j=0}^{d-1} \ket{j} \otimes \ket{\psi_j}$,
where the kets $\ket{\psi_j}$ are unnormalized.  We also take the
convention that
$W \ket{\psi}=\ket{0}$ so that $W^\dagger \ket{0}=\ket{\psi}$.
Using the partition of the state vector, one may show that any
two-qudit state-synthesis operator $W$ can be decomposed 
into  $d-1$ elementary controlled-rotation operators as follows \cite{BOBII}:  
\begin{equation}
\label{eq:divide_ket}
\begin{array}{lll}
W \ &=& \
(V_d \otimes {\bf 1}_d) \;
\prod_{j=0}^{d-2}
\Big[(F_{d-1-j} \otimes {\bf 1}_d)
\wedge_1(V_{d-1-j})\\
& &
(F_{d-1-j}^{\dagger} \otimes {\bf 1}_d)\Big]( {\bf 1}_d \otimes V_0) \, .
\end{array} 
\end{equation}
Here we intend 
$F_{j}=\ket{j}\bra{d-1}+\ket{d-1}\bra{j}+\sum_{k\neq j,d-1} \ket{k}\bra{k}$
to be a state-flip operator.  
The single-qudit operators $V_j$ are chosen so as to perform
$V_j \ket{\psi_j}=t_j^{1/2} \ket{0}$, where $\langle \psi_j | \psi_j \rangle=
t_j$  \cite{selectionQR}.  Then $V_0$ clears the remaining nonzero
amplitudes.

\begin{figure}
\begin{center}
\includegraphics[scale=0.35]{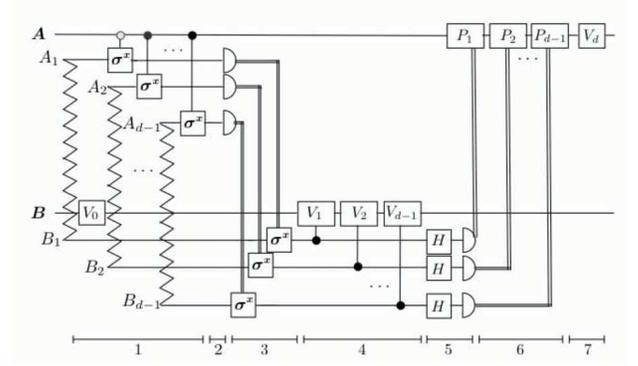}
\caption{\label{fig:2}
A non-local two qudit gate $U=W^\dagger$ 
that realizes the state-synthesis 
$U\ket{0}_{A,B}=\ket{\psi}_{A,B}$ on qudits 
$A$ and $B$ using $d-1$ ancillary qubit pairs (indicated by
sawtooth lines) each prepared in the state 
$\ket{\phi^+}_{A_j,B_j}=1/\sqrt{2}(\ket{00}+\ket{11})_{A_j,B_j}$.  Each 
qubit $A_j(B_j)$in the entangled 
resource can constitute a new particle or a distinct degree of freedom of 
qudit $A(B)$.    Controlled-not gates between $A$ and $A_j$ are conditioned 
on the basis state $\ket{j}_A$, as indicated by the shading of the 
control bubble.  The notations are:  double lines for classical 
controlled operations 
dependent on qubit measurement outcomes, 
$H=e^{i\pi (\sigma^x+\sigma^z)/2\sqrt{2}}$, 
and $P_j=e^{i\pi  \ket{j}\bra{j}}$.  The sequence of 
steps that can be implemented in parallel is indicated at the bottom.}
\end{center}
\end{figure}

The last subsection implicitly describes a non-local implementation
of a controlled (one-qudit state synthesis) operator $W$, 
in that it details a scheme for the non-local
$\wedge_1(V_{d-1-j})$.  The resulting circuit for $W$ is shown in
Fig. \ref{fig:2} and
requires $d-1$ e-bits and $2(d-1)$ c-bits.  Remarkably, the protocol 
can be parallelized to $7$ computational steps.  Here by a single step we mean
a set of operations that is no more time consuming than a controlled 
one qudit rotation 
$\wedge_1(V)$, which itself can be decomposed into controlled-phase 
gates and single 
qudit Givens rotations if so needed.
The only nonobvious parallel step is step $4$. 
Note that the operators $V_j$ generally do not commute.  However, 
just before and just after this step, 
the usual teleportation case study
shows that the state of the system lies within the span of those
$\ket{k}=\ket{k_0}_{A}\otimes_{j=1}^{d-1}\ket{k_j}_{B_j}\otimes \ket{k_d}_B$
in which at most a single $k_j$ is one for $1 \leq j \leq d-1$.
Let $P$ denote the projection of Hilbert space onto the
span of all $\ket{k}$ as above.
If $Q$ denotes the central product of Equation \ref{eq:divide_ket}, we have
\begin{equation}
P  Q P   \  = \ 
\prod_{j=1}^{d-1}e^{-it_j\ket{1}_{B_j}{_{B_j}}\bra{1}\otimes h_j} \\
\end{equation}
For the map of Hamiltonians $h \mapsto P h P$ has image equal to the
span of all $\ket{1}_{B_j}{_{B_j}}\bra{1} \otimes h$.
Moreover, for $j_1\neq j_2$ and any Hermitian $h_1,
h_2$, we have
$\big[ \; \ket{1}_{B_{j_1} B_{j_1}}\bra{1}\otimes h_1,
\ket{1}_{B_{j_2} B_{j_2}}\bra{1} \otimes h_2\; \big]=0$.
Hence we can generate the gates in step $4$ in parallel.
The operations in step $5$ correspond to measurement of qubits $B_j$ in the 
Hadamard basis and count as a single parallel operation.

\subsection{Spectral decomposition bootstrap to nonlocal gates}

This protocol can be extended to implement an arbitrary 
non-local unitary $U\in U(d^2)$ between $A$ and $B$.  
Consider the spectral decomposition Eq.~\eqref{spec} of $U$ 
which involves multiple applications of state-synthesis 
operators $W$ and controlled phase operators $C$.  The 
controlled phase operators are locally equivalent to the operator 
$\wedge_1[{\bf 1}_d+(e^{i\phi}-1)\ket{d-1}\bra{d-1}]$ and thus can be 
implemented in one step using one e-bit and two c-bits. 
Thus, any two-qudit unitary can then be built 
using $\ell = 7\times 2d^2+d^2=15d^2$ parallel operations with the 
assistance of $\#_e=2\times (d-1)\times d^2+d^2=2d^3-d^2$ e-bits and 
$2\#_e$ c-bits. 

Recently, an alternative construction of two-qudit operations using qubit 
entanglement resources was proposed \cite{Zeng:05}.  That work describes 
how a single e-bit and two c-bits suffice to implement a one parameter 
subgroup of $U(d^2)$ between two distant qudits $A$ and $B$ with 
probability one.  Specifically, their protocol realizes unitaries of the 
form $V(\phi)=\exp[i\phi U_A\otimes U_B]$ where the operators $U_A,U_B$ 
are unitary and Hermitian.  However, the authors do not provide an 
algorithm for generating an arbitrary two-qudit unitary nor do they estimate 
the number of e-bits consumed in a covering of $U(d^2)$ with such unitaries.

\subsection{Improved fidelity by purification}

Our protocol requires local high fidelity operations between qudit 
$A$ and a set of qubits $\{A_j\}$ (similarly between $B$ and $\{B_j\}$) 
as well as
high fidelity local unitaries.  In principle, the entangling operations 
might be made
error tolerant.  
Rather than use ancillary qubits that are distinct particles, 
we might use composite
particles endowed a inherent tensor product structure 
$\mathcal{H}=\mathcal{H}_{\mbox{\footnotesize qudit}}\otimes
\mathcal{H}_{\mbox{\footnotesize ancilla}}$ where one subsystem is 
used to encode the qudit and the ancillary subsystem is used to 
assist in two-qudit gate performance.  D\"{u}r and Briegel \cite{Briegel} 
showed that one can perform extremely high-fidelity two {\it qubit} gates with
this partitioning.  In their protocol, information is encoded in one 
two-dimensional degree of 
freedom of each particle, say spin.  Entanglement 
between particles is generated using ancillary degrees of freedom such as
 quantized states of motion along $\hat{x},\hat{y}$ or $\hat{z}$. The prepared 
entanglement may not be perfect.  Yet by using nested entanglement 
purification with two or more degrees of freedom, one can 
prepare a highly entangled state in the ancillary degrees of freedom with 
nonzero probability.  If a purification round fails, then the 
entangled state can be reprepared without disturbing the quantum information 
encoded in the other degree of freedom (here spin).  Given this, a non-local 
${\rm CNOT}$ gate can be implemented between the encoded qubits.

Their protocol is readily extended to non-local gates between qudits using
ancillary qubit degrees of freedom as discussed above.  The 
critical assumption for robustness is that gates which 
couple different degrees 
of freedom of the same particle can be performed with much higher fidelity 
than gates which couple different particles.  
The assumption is frequently valid
because coupling two spatially distinct particles usually involves interactions
mediated by a field which can also couple to the environment and thus 
decohere the system.
In contrast, gates between different degrees of freedom of the same 
particle, such as coupling 
spin to motion in trapped ions \cite{Sackett} or atoms \cite{Haycock} 
can often be implemented with high precision
using coherent control.  

\section{Conclusions}

Quantum computation with qudits requires more control at the single particle
level than with qubits.  It might be expected that the additional 
time needed to control all the levels would be prohibitively long 
in terms of memory decoherence times.  We have shown how parallel (time-step optimized)
one-qudit and two-qudit computation help surmount such difficulties.
Given a qudit with a connected
coupling graph, the time complexity for constructing an arbitrary unitary
can be reduced at the expense of additional control resources.  Even 
for systems with little connectivity between states, 
such as in the case of a qudit encoded in hyperfine levels of an 
atomic alkali, the number of parallel elementary gates can be made 
close to the optimal count for a maximally connected 
state space.  For the purposes of two-qudit gates, we found a 
non-local implementation of 
an arbitrary unitary using $O(d^2)$ parallel steps.  The protocol 
uses $O(d^3)$ e-bits which could be in principle be prepared and 
distributed ahead of time with high fidelity.

Some outstanding issues remain.  First, our treatment focused on systems with allowed 
couplings between pairs of states.  In other systems, the selection rules may 
dictate a 
different set of subalgebras to be used for quantum control, e.g. 
spin-$j$ representations 
of the algebra $\mathfrak{su}(2)$.  Some particular computations 
may be realized with
much greater efficiency using such generators.  Second, fault
tolerant computation relies not on exactly universal computation, but rather 
by approximating unitaries using a discrete set of one and two-qudit gates.
It would be worthwhile to investigate optimal protocols for implementing a discrete set 
of fault tolerant non-local two qudit gates using entangled qubit pairs.
 
\acknowledgments
The work of DPO was supported in part
by the National Science Foundation under Grants CCR-0204084 and CCF-0514213.  
GKB received support from an DARPA/QUIST grant.
We are grateful to Samir Khuller for providing reference \cite{heyesha}.


\begin{thebibliography}{99}
\bibitem{footnote1}
{Note that this transition corresponds to a 
$\ket{F_{\downarrow},M_F=0}\rightarrow \ket{F_{\uparrow},M_F=0}$.  
Selection rules for Raman laser pulses only allow this transition if 
both the pump and probe pulses are $\sigma_{+}$ or $\sigma_{-}$ 
polarized.  Because this particular  transition is not needed for 
parallelism we can fix the pump or probe pulse to have $\pi$ polarization 
and still access all the other couplings in the graph using 
frequency selectivity.}
\end{thebibliography}
\end{document}